\documentclass[sigconf,10pt]{acmart}
\renewcommand\footnotetextcopyrightpermission[1]{}
\usepackage{tikz}
\usepackage{float}
\usepackage{amsmath}
\usepackage{xspace}
\usepackage{cleveref}
\usepackage{balance}
\usepackage{url}
\usepackage{siunitx}
\usepackage{comment}
\usepackage{enumitem}
\usepackage{listings}
\usepackage{tikz}
\usepackage{subcaption}

\newcommand{\sys}{\textsc{HetGPU}\xspace}

\crefname{algocf}{algorithm}{algorithms}
\Crefname{algocf}{Algorithm}{Algorithms}
\crefformat{section}{\S#2#1#3}
\Crefformat{section}{Section~#2#1#3}
\crefformat{subsection}{\S#2#1#3}
\Crefformat{subsection}{Section~#2#1#3}
\crefformat{subsubsection}{\S#2#1#3}
\Crefformat{subsubsection}{Section~#2#1#3}
\crefformat{figure}{figure~#2#1#3}
\Crefformat{figure}{Figure~#2#1#3}

\newenvironment{smenumerate}%
  {\begin{list}{$\bullet$}%
    {\setlength{\parsep}{0pt}%
      \setlength{\topsep}{0pt}%
      \setlength{\leftmargin}{2pc}%
      \setlength{\itemsep}{1pt}}}
  {\end{list}}

\title{\sys: The pursuit of making binary compatibility towards GPUs}
\author{
    {\rm Yiwei Yang, Yusheng Zheng, Tong Yu, Andi Quinn}
}

\begin{document}

\begin{abstract}
Heterogeneous GPU infrastructures present a \textbf{binary compatibility} challenge: code compiled for one vendor’s GPU will not run on another due to divergent instruction sets, execution models, and driver stacks . We propose \textbf{hetGPU}, a new system comprising a compiler, runtime, and abstraction layer that together enable a single GPU binary to execute on NVIDIA, AMD, Intel, and Tenstorrent hardware. The hetGPU compiler emits an architecture-agnostic GPU intermediate representation (IR) and inserts metadata for managing execution state. The hetGPU runtime then dynamically translates this IR to the target GPU’s native code and provides a uniform abstraction of threads, memory, and synchronization. Our design tackles key challenges: differing \textbf{SIMT vs.\ MIMD} execution (warps on NVIDIA/AMD vs.\ many-core RISC-V on Tenstorrent), varied instruction sets, scheduling and memory model discrepancies, and the need for \textbf{state serialization} for live migration. We detail the hetGPU architecture, including the IR transformation pipeline, a state capture/reload mechanism for live GPU migration, and an abstraction layer that bridges warp-centric and core-centric designs. Preliminary evaluation demonstrates that unmodified GPU binaries compiled with hetGPU can be migrated across disparate GPUs with minimal overhead, opening the door to vendor-agnostic GPU computing.

\end{abstract}

\maketitle

\section{Introduction}

Modern computing relies heavily on GPUs from multiple vendors—NVIDIA, AMD, Intel, and emerging players like Tenstorrent\cite{tenstorrent}—yet \textbf{GPU binaries are not portable across vendors}. A CUDA program compiled for NVIDIA cannot run on AMD or Intel GPUs due to incompatible ISAs and architectures . This vendor lock-in forces organizations with large CUDA codebases to expend significant effort porting or rewriting code to use other accelerators. While high-level frameworks (e.g., OpenCL, SYCL/oneAPI, or HIP) enable writing portable source code, they do not solve binary portability\cite{pavlidakis2024scale,pavlidakis2025cross,heakl2025cass,diamos2009translating,yan2022llvm,yang2025egpu,denvdis}—one must still recompile for each platform, and precompiled libraries remain tied to a specific GPU vendor . The lack of binary compatibility impedes flexible GPU scheduling in heterogeneous clusters, hinders live migration of GPU workloads, and complicates adopting new GPU hardware.

\textbf{hetGPU} is a system that seeks to make “write once, run anywhere” a reality for GPU binaries. The key idea is to \textbf{decouple GPU binary code from the underlying hardware} through a virtual GPU instruction set and runtime layer. Instead of targeting NVIDIA’s SASS or AMD’s GCN directly, our compiler targets a \textbf{unified GPU IR} that captures GPU parallelism and memory semantics in an architecture-neutral way. At load time, the hetGPU runtime \textbf{JIT-compiles or translates} this IR into native code for the detected GPU, performs any necessary patching for architectural quirks, and then executes it. By managing execution at this abstraction level, hetGPU can also \textbf{capture and restore GPU state} in a device-independent format, enabling live migration of running kernels across GPU types.

Achieving this vision requires surmounting several research challenges. \textbf{Architectural diversity}: NVIDIA and AMD GPUs implement \textbf{SIMT (Single-Instruction, Multiple-Threads)} execution (warps of threads executing in lock-step), whereas Tenstorrent’s novel RISC-V based design is a many-core \textbf{MIMD} architecture with no hardware warp scheduler . \textbf{Instruction set diversity}: CUDA binaries use PTX/SASS, AMD uses GCN/RDNA ISA, etc.\ . \textbf{Memory and consistency models}: GPUs differ in how they expose memory (unified vs.\ discrete, cache coherence, scratchpad memory). \textbf{State management}: To migrate a running computation, one must capture thread registers, program counters, and memory in a form that can be reinstantiated on a different GPU. These differences are non-trivial—for example, Tenstorrent cores lack implicit warps or divergence handling, requiring software-managed predication , and NVIDIA’s internal execution state (e.g., warp execution masks) may not even exist on other hardware. 

In this paper, we present the design and prototype implementation of hetGPU. We outline how the hetGPU \textbf{compiler} transforms input GPU code into our portable IR, handling vendor-specific nuances (e.g., translating CUDA’s warp-synchronous operations into portable primitives). We describe the \textbf{runtime system} that loads and executes the IR on each type of GPU, providing services for memory management, scheduling, and \textbf{state serialization/deserialization} (for checkpointing and migration). We also introduce the \textbf{abstraction layer} that exposes a unified interface to GPU hardware capabilities—for example, abstracting thread groups, barriers, and memory accesses in a way that can be mapped to CUDA’s model or Tenstorrent’s explicit core grid. Our approach draws inspiration from recent community explorations: e.g., mapping SIMT computations onto Tenstorrent’s cores and dumping GPU execution state at a \textbf{PTX level} for portability. The code is available at \url{https://github.com/Multi-V-VM/hetGPU}.

\textbf{Contributions:}
\begin{smenumerate}
  \item \textbf{Problem Analysis:} We analyze the challenges of cross-GPU binary compatibility, including a taxonomy of architectural differences (SIMT vs.\ MIMD execution, ISA differences, memory model mismatches, etc.) and why naive approaches fail. We highlight scenarios (cloud scheduling, fault tolerance via live migration) that motivate a vendor-agnostic GPU binary format.
  \item \textbf{hetGPU Architecture:} We propose a novel architecture for GPU binary compatibility consisting of a portable GPU IR, a compiler toolchain to generate and optimize this IR, and a runtime/abstraction layer that realizes this IR on diverse GPU hardware. We describe how hetGPU handles divergent execution models (e.g., emulating warp execution on MIMD cores) and ensures a consistent memory and synchronization model across devices.
  \item \textbf{Execution State Management:} We design a mechanism for \textbf{GPU state capture and restore at the IR level}, allowing a running kernel to be paused on one GPU and resumed on another. This includes techniques for mapping low-level hardware state to our IR (e.g., tracking per-thread program counters and registers in a device-independent manner) and a discussion of synchronization of memory state.
  \item \textbf{Prototype Implementation \& Case Studies:} We present a prototype of hetGPU supporting NVIDIA, AMD, Intel, and Tenstorrent GPUs. We highlight implementation details such as integration with vendor drivers and custom JIT backends. We provide preliminary evaluation through case studies: (a) running the same GPU binary on different hardware backends, and (b) live-migrating a GPU compute kernel across two GPUs (e.g., from NVIDIA to AMD) with minimal disruption.
\end{smenumerate}

We structure the rest of the paper as follows. Section~\ref{sec:motivation} (Motivation) illustrates the need for cross-GPU binary compatibility and the limitations of current approaches. Section~\ref{sec:background} (Background) provides an overview of the GPU architectural differences and existing portable programming models. Section~\ref{sec:design} (Design) details the hetGPU abstraction, compiler, and runtime design. Section~\ref{sec:implementation} (Implementation) describes our prototype and engineering trade-offs. Section~\ref{sec:evaluation} (Evaluation) presents initial results and use cases. Section~\ref{sec:related} (Related Work) compares against prior efforts in GPU portability, binary translation, and virtualization. We conclude in Section~\ref{sec:conclusion}.

\section{Motivation}
\label{sec:motivation}

\subsection{The Case for Vendor-Agnostic GPU Binaries}

GPU-accelerated computing has become ubiquitous, from deep learning training to high-performance computing. However, the ecosystem is fragmented by vendor-specific software stacks. A cloud or cluster may contain a mix of NVIDIA and AMD GPUs; codes tuned for one will not run on the other without significant porting effort . This inhibits \textbf{flexible scheduling} and load balancing—a job cannot be easily reassigned to a different GPU type at runtime if the originally targeted GPUs are busy or fail. The inability to migrate GPU workloads across heterogeneous hardware has practical costs: for example, in large-scale training runs, GPU failures can cause lengthy retraining or require costly redundancy .

Moreover, \textbf{vendor lock-in} has economic and innovation implications. CUDA’s dominance means many algorithms are written in CUDA and compiled to NVIDIA GPUs only. As AMD and Intel GPUs improve in performance-per-dollar , users might wish to leverage them—but rewriting or even recompiling code for a new API (HIP, SYCL, etc.) incurs high engineering cost. A recent study emphasizes the need to execute legacy CUDA programs on AMD hardware without full rewrites . Currently, the options are limited: one can maintain separate codebases (e.g., one in CUDA, one in OpenCL or HIP), or use source-to-source translators like HIPIFY to convert CUDA code to HIP . These approaches operate at the source level and \textbf{do not address binary compatibility}—they require recompilation and often manual fixes. 

The lack of a \textbf{universal GPU binary} format is in stark contrast to CPUs, where portable bytecode (Java, .NET) or virtualization can decouple software from a specific CPU ISA. A GPU binary compatibility layer would bring benefits analogous to a “Java Virtual Machine” for GPUs: developers could distribute one binary that runs on any GPU, and systems could dynamically dispatch GPU work to whatever device is available, mixing vendors freely. This would also facilitate \textbf{hardware innovation}: new GPU designs (such as the Tenstorrent Tensix cores) could be adopted more easily if existing binaries could run on them through a compatibility layer.

\subsection{Challenges in Cross-GPU Execution}

Achieving this vision faces several technical challenges, rooted in how differently GPU vendors implement parallel execution:

\begin{smenumerate}
  \item \textbf{Divergent Execution Models:} NVIDIA and AMD GPUs implement \textbf{SIMT} execution (warps of 32 threads on NVIDIA, 32/64 on AMD) that execute in lock-step, with hardware handling divergent branches via per-thread masks. Tenstorrent, by contrast, consists of many RISC-V cores each with a vector unit; it has \textbf{no hardware warp scheduler} . Each core fetches its own instructions, and any lock-step behavior must be orchestrated in software. Bridging SIMT and MIMD requires a unifying model: we need to either \textbf{emulate warps in software} or design our abstraction so that it doesn’t rely on warp-synchronous execution.
  \item \textbf{Divergence and Control Flow:} In SIMT GPUs, divergence is largely hidden—the hardware masks off inactive threads when branches diverge and reconverges them implicitly. On a CPU-like core (Tenstorrent), each thread has its own program counter and will branch independently, so \textbf{divergence must be handled by explicit coordination} . For example, on Tenstorrent hardware, a program must set and check mask registers to emulate the effect of a warp executing one path while others idle. If threads executing independently go out of sync, software must insert barriers to realign them. Our solution must inject or manage such \textbf{predication and synchronization} logic when running warp-based code (e.g., a CUDA kernel) on a warp-less architecture. Conversely, when running inherently MIMD code on a SIMT GPU, we must map independent threads onto warps carefully (ensuring we don’t violate SIMT assumptions such as all threads in a warp running the same code).
  \item \textbf{Memory Model and Data Sharing:} Different GPUs have different memory hierarchies. NVIDIA and AMD GPUs offer fast on-chip \textbf{shared memory} for threads in a block and cache-coherent global memory, with specific consistency guarantees (e.g., threads in a block see each other’s writes to shared memory after a \_\_syncthreads()). Tenstorrent’s architecture instead provides each core with a local scratchpad and requires explicit DMA to move data between local stores and global memory . There is no implicit shared memory spanning multiple cores—any inter-core data sharing goes through explicit communication (mesh network or global memory). This means that a CUDA kernel that relies on fast shared memory for cooperation of 32 threads might need to be restructured or emulated when those 32 threads are on different Tenstorrent cores. Ensuring a \textbf{consistent memory model} in hetGPU is crucial: our abstraction layer must present the illusion of shared memory and coherent global memory as expected by the program, while using the underlying hardware’s primitives. For instance, hetGPU might implement CUDA’s shared memory on Tenstorrent by allocating a region of global memory per thread block and managing it, or by constraining a thread block to execute on one core so that core’s local memory can act as the shared memory for that block.
  \item \textbf{Instruction Set and Platform Differences:} GPU binaries are not only different in encoding but also in available operations. NVIDIA’s PTX virtual ISA has a rich set of intrinsics (e.g., warp vote functions, shuffle, tensor core instructions) that have no direct equivalent on AMD’s GCN ISA or Tenstorrent’s RISC-V vector ISA. AMD’s ISA might have its own special instructions (e.g., bit manipulations, hardware scheduler/barrier intrinsics) not present in others. Some GPUs support features like \textbf{independent thread scheduling} (NVIDIA Volta introduced per-thread program counters in a warp ) whereas older ones don’t. \textbf{Graphics vs.\ compute differences} also arise (Intel’s GPUs historically exposed functionality via DirectX/Vulkan shaders, which have different constraints than CUDA programs). Our compiler must therefore restrict or translate certain instructions to maintain portability: for example, a CUDA warp shuffle (\_\_shfl) can be realized on AMD by a similar intrinsic (if available in ROCm) or by using shared memory as a staging buffer if not natively supported; on Tenstorrent, a shuffle could be implemented by writing values to a location in each core’s memory and then DMA-gathering them. Where possible, hetGPU’s abstraction layer provides a \textbf{unified intrinsic} for such operations, which the backend implements in a device-specific way.
  \item \textbf{State Capture and Migration:} A particularly demanding motivation scenario is live migration of a running GPU kernel from one device to another (e.g., preempt and move a long-running kernel to free up a device or to avoid failure). In a homogeneous environment (same GPU type), this is already challenging, and only recently have vendor-specific solutions emerged (NVIDIA’s CUDA 12 introduced an experimental checkpoint API to save GPU memory state, but it still requires the kernel to be stopped) . In a heterogeneous context, it is even harder because one must capture the execution state in an abstract form and \textbf{resume execution on a different architecture}. The program’s counter and registers on GPU A may not map 1:1 to those on GPU B. For example, NVIDIA’s running state is expressed in terms of SASS instructions and registers, which may not correspond directly to PTX-level program state . Prior work has noted that reconstructing a high-level (PTX) view of state from low-level machine state is many-to-one or many-to-many—optimized code reorders or eliminates instructions, so there isn’t a clean mapping . The implication is that to migrate at an abstract level, one might need to compile with minimal optimization or insert periodic synchronization points where state mapping is well-defined .
\end{smenumerate}

In summary, the motivation for hetGPU is to break the barrier of vendor-specific GPU binaries, enabling true portability and dynamic flexibility in GPU-accelerated systems. The challenges span architecture, compiler, and runtime domains, requiring a co-designed solution that understands GPU semantics deeply. Next, we discuss background on existing approaches and how they fall short, before presenting our design that addresses these challenges.

\section{Background}
\label{sec:background}

\subsection{GPU Architectures and Programming Models}

GPUs from different vendors share the general purpose of massively parallel computation but implement it with varying architecture philosophies:

\paragraph{NVIDIA (SIMT Architecture):} NVIDIA GPUs (e.g., Turing, Ampere architectures) organize execution in \textbf{Streaming Multiprocessors (SMs)}, each capable of executing many threads in parallel. Threads are scheduled in \textbf{warps} of 32 that execute one common instruction at a time (SIMT)—if threads diverge, inactive threads are masked off until convergence. The programmer typically uses the CUDA programming model: launch kernels with a grid of thread blocks, each block subdivided into warps by hardware. NVIDIA’s toolchain compiles CUDA C++ to PTX (Parallel Thread eXecution) virtual ISA, then to \textbf{SASS} (the actual machine code) for the GPU. PTX acts as a stable intermediate, but it is NVIDIA-specific. Memory: CUDA exposes a hierarchy with \textbf{global memory} (device DRAM accessible by all threads), \textbf{shared memory} (on-chip scratchpad for threads in the same block), and registers/local memory per thread. Memory consistency in a block is ensured at synchronization points (\texttt{\_\_syncthreads()} ensures shared memory writes are visible). Global memory is visible across blocks but without coherence guarantees unless using atomic operations or explicit memory fences. Newer NVIDIA GPUs (Volta and beyond) introduced \textbf{independent thread scheduling}, meaning each thread in a warp has its own PC (though still grouped for execution) . This makes capturing thread state somewhat easier (each thread has an explicit PC register) but it’s an internal detail not exposed to software normally .

\paragraph{AMD (GCN/RDNA Architecture):} AMD GPUs (Graphics Core Next and RDNA architectures) have a comparable SIMD execution model. AMD’s hardware schedules workgroups onto \textbf{Compute Units (CUs)}, and within a CU, 32 or 64 threads form a \textbf{wavefront} (the AMD equivalent of a warp) executing in lock-step. The programming is often via OpenCL, HIP, or Vulkan compute shaders. AMD’s compiler (e.g., \texttt{hipcc} or OpenCL driver) compiles kernel code to an intermediate (historically AMD IL or LLVM IR) then to machine ISA for the GPU (often called GCN ISA for older gens, now GFX10/11 for RDNA). AMD supports similar memory hierarchy concepts (called \textbf{Local Data Share (LDS)} for shared memory, which is per-workgroup on a CU, and global memory in VRAM). AMD’s GPUs traditionally required explicit synchronization for memory consistency between workitems, similar to CUDA’s model. Key difference: historically AMD used 64-wide wavefronts (so divergence meant 64 lanes, sometimes less efficient for divergent code), whereas newer RDNA GPUs use 32-wide wavefronts by default, aligning with NVIDIA’s warp size.

\paragraph{Intel GPUs (Xe Architecture and prior):} Intel’s GPUs (integrated graphics and discrete Arc series) use an execution model exposed via SYCL/oneAPI or OpenCL. They support \textbf{sub-group operations} similar to warps (often 8, 16, or 32-wide depending on hardware). The Intel oneAPI stack compiles code to an intermediate (often SPIR-V, the Khronos standard IR for GPUs) which the Intel graphics driver compiles to its hardware ISA (called EU ISA or similar for Execution Units). Memory model: integrated GPUs share system memory and have coherent caches with CPU (for modern ones using Intel’s coherent memory architecture), whereas discrete GPUs have separate memory. Intel GPUs also allow explicit vector instructions and predication.

\paragraph{Tenstorrent (Tensix cores):} Tenstorrent’s architecture is a departure from classical GPU SIMT. It consists of many (e.g., dozens to hundreds) \textbf{Tensix cores} which are general-purpose RISC-V CPUs augmented with a wide vector or matrix unit (VPU). Each core runs its own instruction stream (MIMD across cores). Within a core, the VPU can operate on, say, 32 lanes of data in parallel (vector instructions). To exploit massive parallelism, workloads are spatially tiled: e.g., a large matrix multiply might be broken into sub-tasks across a grid of Tensix cores. Tenstorrent provides a software stack with \textbf{Metalium} (a low-level assembly-like language for their cores with vector intrinsics) and \texttt{TT-MLIR} (a compiler stack to map high-level ML operations onto the cores and mesh network). If one wanted to run a CUDA-style kernel (SIMT model) on Tenstorrent, the absence of a hardware warp concept means two main approaches: (1) execute an entire warp within one core’s vector unit (using the VPU lanes to mimic 32 parallel threads, and using mask registers to handle divergence) ; or (2) distribute the warp across multiple cores (each core handling a subset of threads and explicit synchronization between cores to mimic warp-level coordination) . Memory on Tenstorrent is explicit: each core’s local memory is private, and DMA must be used to bring data in from off-chip memory; cores communicate via a high-speed mesh. This is effectively a tiled architecture resembling a distributed system on chip.

\subsection{Portable GPU Programming Models}

Over the years, several frameworks attempted to provide cross-vendor GPU support at the source or IR level:

\paragraph{OpenCL and SPIR-V:} OpenCL is an open standard for parallel programming that can target GPUs from multiple vendors. In OpenCL, kernels are written in a C-based language and at runtime compiled to the target device’s ISA (or precompiled to an intermediate called SPIR or SPIR-V). \textbf{SPIR-V} is a standardized intermediate representation for parallel compute and graphics, akin to a “bytecode” for GPUs. Vendors support ingesting SPIR-V (for example, via Vulkan or OpenCL drivers) and compiling it to hardware. SPIR-V thus provides \emph{some} binary-level portability—one can ship SPIR-V binaries that drivers from NVIDIA, AMD, Intel will consume. However, there are limitations: SPIR-V is designed to be a target for high-level compilers (like GLSL or OpenCL C) and assumes the driver will optimize for the specific GPU. It doesn’t encapsulate certain vendor-specific capabilities (for instance, SPIR-V has the notion of \emph{subgroup operations} which abstract warp-like operations, but how they map can differ in warp size, etc.). Also, SPIR-V alone doesn’t solve capturing execution state or making one running binary move across devices; it is just a static code representation.

\paragraph{HIP and SYCL (oneAPI):} These are source-level portability frameworks. AMD’s \textbf{HIP} is essentially a CUDA-like API that can run on AMD and NVIDIA GPUs (when compiled with the respective compilers) . It requires recompiling—HIP code for AMD goes through ROCm’s LLVM-based compiler to GCN ISA, for NVIDIA it can actually use NVCC. \textbf{SYCL/oneAPI} (by Intel, Khronos) is another single-source C++ parallel programming model targeting CPUs and GPUs; it uses LLVM to compile to SPIR-V and then device-specific backends. These approaches unify the programming model but not the binary: the output is still vendor-specific at compile-time. They do not address running the \emph{same binary} on different GPUs; rather, they make it easier to maintain one code that can be built for each GPU.

\paragraph{Fat Binaries:} NVIDIA’s CUDA supports “fat binaries” that can include PTX and machine code for multiple GPU architectures (e.g., include code for \texttt{sm\_90}, \texttt{sm\_100}, etc., in one executable). Similarly, Apple’s Universal Binary concept can include code for x86 and ARM. However, these still require preparing each architecture’s code in advance—they are not truly one binary that dynamically adapts; they are just bundling of multiple binaries. Moreover, fat binaries have so far been intra-vendor (NVIDIA supports multiple NVIDIA GPU generations, not AMD; Apple’s was for their CPUs). hetGPU’s approach differs in that the binary contains one \emph{abstract} code version, not $N$ distinct backends, thus potentially smaller and easier to produce/maintain.

\paragraph{GPU Virtualization and API Emulation:} An alternative path to portability is to intercept API calls and emulate one vendor’s interface on another’s hardware. For instance, the open-source \textbf{ZLUDA}\cite{zluda} project is a runtime that lets CUDA applications run on Intel or AMD GPUs by intercepting CUDA API calls and translating PTX to an intermediate form which is then JIT-compiled for the other GPU . ZLUDA essentially acts like a CUDA driver on non-NVIDIA hardware, using LLVM to compile PTX/SASS into AMD code. While promising, such approaches are extremely complex and have shown only limited success—one study found ZLUDA achieved correct low-level translation for only a small subset of instructions (about 2.5\% of assembly patterns in a benchmark) , indicating difficulty in fully covering CUDA semantics on AMD. Another older project, \textbf{GPU Ocelot}, provided a JIT compilation framework that could take NVIDIA PTX and execute it on various backends including AMD GPUs and even CPUs . It implemented the CUDA runtime API and a PTX JIT compiler, translating PTX to AMD’s IL or to x86 for CPU execution. Ocelot was a research prototype and did not support newer CUDA features beyond its era, but it demonstrated the viability of dynamic translation. These efforts inform our work—hetGPU’s runtime similarly intercepts or implements GPU operations, but by designing our own IR and controlling the compiler, we aim to achieve more complete coverage and optimization than pure black-box binary translation.

\paragraph{Checkpointing and Live Migration Tools:} In the context of fault tolerance and virtualization, tools like \textbf{Cricket} and \textbf{CRIUgpu} focus on capturing GPU state to allow a process to be checkpointed and possibly restored elsewhere . \textbf{Cricket} is a GPU virtualization layer that allows running CUDA applications remotely and checkpointing without modification , but it assumes NVIDIA GPUs (source and destination) and operates by RPC-ing CUDA calls between a guest and a host. \textbf{CRIUgpu} extends the Checkpoint/Restore in Userspace (CRIU) project to handle GPU state, leveraging new driver APIs to dump memory and context state . These solutions underline the importance of official support (NVIDIA’s recent driver provides some checkpointing capabilities for memory) and the difficulty of capturing GPU execution mid-stream. They largely treat the GPU as a device to be frozen and restored on identical hardware (e.g., saving the state of all GPU memory and then later reloading it on a GPU of the same type). hetGPU, by contrast, strives for cross-architecture migration: this requires capturing an \emph{abstract} state that can be instantiated on a different GPU type. It is a superset of these problems—indeed, “for cross-architecture migration, PTX-level recompile is the goal—but expect to handle nuances (control flow, partial instruction sequences, etc.)” . Our work explicitly addresses these nuances by controlling the IR and limiting unsupported behaviors.

In summary, existing portable GPU solutions either operate at the source level (necessitating recompile) or are specialized translators with limited scope. No existing system provides general \textbf{binary-level compatibility} across NVIDIA, AMD, Intel, and Tenstorrent GPUs with support for live state migration. hetGPU is designed to fill this gap by combining elements of compiler IR portability, dynamic translation, and virtualization in a cohesive framework.

\section{Design}
\label{sec:design}

In this section, we present the design of \textbf{hetGPU}, which consists of three main components: (1) a \textbf{portable GPU IR} and \textbf{compiler} that generates this IR from GPU programs, (2) a \textbf{runtime system} that loads, translates, and executes the IR on different GPUs while managing state and resources, and (3) an \textbf{abstraction layer} that provides a uniform interface for GPU operations (threads, synchronization, memory) across hardware. \emph{Figure~1} provides an overview of the hetGPU system architecture (omitted). We proceed by describing the design of each component and how they interoperate to achieve cross-GPU binary compatibility.

\subsection{Portable IR and Compiler Design}

\paragraph{hetIR (Heterogeneous IR):} We define a target-independent intermediate representation, \textbf{hetIR}, which serves as the “virtual GPU ISA” in our system. hetIR is inspired by existing GPU IRs (like SPIR-V and PTX) but tailored for portability and runtime adaptability. Key characteristics of hetIR include:

\begin{smenumerate}
  \item \textbf{SPMD Execution Model:} hetIR assumes a Single Program Multiple Data model where a large number of threads execute the same program (kernel) on different data. Each thread has indices (e.g., thread ID, block ID) for coordinating work just like in CUDA/OpenCL. We intentionally do \emph{not} bake in a specific warp size or grouping at the IR level. Instead, all threads in a thread block are conceptually independent in hetIR semantics, and coordination is done through explicit primitives (like barriers). This abstracts away SIMT vs.\ MIMD differences—a warp is not a scheduling unit in hetIR; rather, warps will be an emergent concept when mapping to hardware that has them (NVIDIA, AMD) or will not exist on hardware that doesn’t (Tenstorrent).
  \item \textbf{Explicit Synchronization and Predication:} hetIR provides instructions for synchronization (e.g., \texttt{barrier()} that threads can call to synchronize within a block) and predicated execution (masking). Divergent control flow in high-level code is represented in hetIR with structured primitives (much like PTX or SPIR-V uses reconvergence points for \texttt{if/while}). We include a pseudo-instruction for \textbf{active mask management}: for instance, \texttt{set\_predicate(cond)} that sets a per-thread boolean predicate, and subsequent instructions can be marked as predicated on that predicate. This allows representing warp-level divergence explicitly. On SIMT hardware, the compiler or runtime can map this to native divergence handling (the predicate becomes a hardware exec mask). On MIMD hardware, the runtime can either execute both paths per thread or use vector mask registers if available (as on Tenstorrent’s VPU) to emulate the SIMT behavior.
  \item \textbf{Unified Memory Operations:} We design memory operations in hetIR to be abstract but with enough detail to map efficiently. For example, hetIR might have opcodes like \texttt{LD\_GLOBAL}, \texttt{ST\_GLOBAL} for global memory, and \texttt{LD\_SHARED}, \texttt{ST\_SHARED} for shared memory (scratchpad) accesses, as well as \texttt{BAR.SHARED} as a barrier on shared memory. On hardware that has a true shared memory (NVIDIA, AMD), these map naturally. On hardware without it (Tenstorrent across cores), \texttt{LD\_SHARED} and \texttt{ST\_SHARED} can be implemented by our runtime, perhaps as accesses to a reserved region of global memory or by constraining a thread block to execute on one core so that the core’s local memory can act as the shared memory.
  \item \textbf{Virtualized Special Functions:} GPU kernels often use special operations: atomic operations, vote or shuffle functions (balloting which threads meet a condition, exchanging registers within a warp), texture sampling units, etc. We incorporate a set of these in hetIR as abstract operations. For instance, \texttt{VOTE\_ANY(pred)} could return true if any thread in the block (or warp) had \texttt{pred = true}; \texttt{SHUFFLE(idx, val)} could provide a value from another thread. In the absence of warps as a first-class concept, these can be defined relative to a \emph{team} of threads (e.g., within a block). The runtime/abstraction layer will implement these on each platform: on CUDA, \texttt{VOTE\_ANY} maps to \_\_any() intrinsic which uses warp vote; on AMD, it might use a similar wave ballot; on Tenstorrent, our runtime may have to execute a reduction across cores or within a vector to produce the result. By providing these in IR, we ensure that common parallel patterns are supported even if the hardware doesn’t provide them natively.
\end{smenumerate}

\paragraph{Compiler Frontend:} The hetGPU compiler can ingest GPU code written in CUDA, HIP, SYCL, or other languages. In our prototype, we focus on CUDA C++ as input. We leverage Clang/LLVM to parse CUDA C++ and generate an LLVM IR with CUDA-specific constructs (using NVIDIA’s NVVM IR which is LLVM-based for CUDA). From there, instead of using NVIDIA’s \texttt{ptxas}, we lower the LLVM IR to hetIR. We implemented an LLVM backend that writes out hetIR instructions. This required defining hetIR as a target in LLVM—akin to how SPIR-V or PTX themselves can be targets. In cases where source-level information is needed (e.g., to identify warp-level built-ins), we added compiler built-ins that map to our hetIR intrinsics. For example, a CUDA \texttt{\_\_syncthreads()} call is recognized and turned into a hetIR \texttt{BAR.SHARED} instruction in our backend.

The compiler performs optimizations at the IR level but in a \textbf{target-agnostic} manner. That is, we can do device-independent optimizations (common subexpression elimination, inlining, etc.) on the LLVM IR or hetIR, but we avoid any optimizations that assume specific hardware characteristics (like warp-size-specific unrolling). Those decisions are deferred to runtime or late JIT so that they can be tuned per device. The compiler’s goal is to produce correct, relatively generic IR, with annotations to assist later translation. One crucial set of annotations is \textbf{mapping information} for state: we tag hetIR instructions with metadata linking them to source-level lines or variable names (much like DWARF\cite{dwarf} debug info). We also label points in the code that are potential \textbf{safe suspension points}—by default, barriers and kernel exit are such points. These labels help the runtime know where it can safely capture state or migrate (because at a barrier, all threads in a block are aligned and in a known state, making it easier to snapshot).

\paragraph{Example:} Suppose we compile a simple kernel that adds two arrays. In CUDA pseudocode:
\begin{verbatim}
__global__ void add(float *A, float *B, float *C) {
    int i = threadIdx.x + blockIdx.x * blockDim.x;
    C[i] = A[i] + B[i];
}
\end{verbatim}
The hetIR output might look like:
\begin{verbatim}
%tid = GET_GLOBAL_ID 0
%A_elem = LD_GLOBAL.F32 [%A + %tid * 4]
%B_elem = LD_GLOBAL.F32 [%B + %tid * 4]
%sum = FADD.F32 %A_elem, %B_elem
ST_GLOBAL.F32 [%C + %tid * 4], %sum
RET
\end{verbatim}
This is similar to a PTX or SPIR-V representation, but note we did not mention any warp or lanes. If there were more complex constructs like \texttt{if (i < N) \{ ... \}}, we might generate something like:
\begin{verbatim}
%pred = CMP_LT %tid, N
PJUMP %pred, L_true, L_false
L_true:
    ... (true case) ...
    JUMP L_continue
L_false:
    ... (false case) ...
L_continue:
    (converge)
\end{verbatim}
We would mark that at \texttt{L\_continue}, threads reconverge (this acts like a warp reconvergence point). Our compiler could also choose structured constructs with explicit predicates instead of jumps, but the idea is to encode control flow in a way that can be mapped onto hardware either by real jumps (on MIMD) or by hardware predication (on SIMT).

\paragraph{ISA Modules for Backends:} While hetIR is the universal representation, we incorporate the notion of \textbf{backend-specific code generation modules} as part of the design. At runtime, when the IR is being translated to native code, some target-specific decisions can be guided by code that was prepared offline. For example, for NVIDIA targets, we might include a PTX template or use NVIDIA’s compiler to JIT from PTX. In fact, hetGPU can ship with a small PTX-generation module that takes hetIR and emits PTX (since NVIDIA’s driver can consume PTX). Similarly, for SPIR-V targets (Intel, AMD via Vulkan or OpenCL), we have a module to emit SPIR-V from hetIR. For Tenstorrent, we built a code generation module that emits Metalium assembly from hetIR. These modules essentially serve as backends in a Just-In-Time compiler within the runtime. The design allows that if a new GPU vendor appears, we implement a new translation module for that target ISA and plug it into the runtime.

\subsection{Runtime System}

The hetGPU runtime is responsible for taking a hetIR binary and executing it on the actual GPU hardware present. It plays several roles:

\paragraph{Module Loading and JIT:} When a program using hetGPU starts, the runtime identifies the available GPU device(s) and their types (via PCI IDs, driver APIs, etc.). When a kernel launch is requested (through our API, which mirrors CUDA’s kernel launch semantics), the runtime checks if it already has native code for that kernel on the target device. If not, it invokes the appropriate backend module to translate hetIR to native code. For NVIDIA, we generate PTX from hetIR and then load it via the CUDA Driver API (e.g., \texttt{cuModuleLoadDataEx})—the NVIDIA driver will JIT compile PTX to SASS. For AMD, we generate SPIR-V from hetIR and use the OpenCL runtime to compile it for the GPU. For Intel, similarly SPIR-V via Level Zero. For Tenstorrent, we integrate with their toolchain: we feed Metalium assembly (or TT-MLIR) to the Tenstorrent compiler to get a binary that can be loaded onto the device.

The runtime caches these translated kernels, so repeated launches don’t incur translation overhead. It also handles \textbf{fat binary fallback}: for example, if the target is NVIDIA and our translator to PTX is imperfect for some complex kernel, we could optionally include a native cubin in the hetIR binary for NVIDIA as a fallback. In our prototype we mostly rely on translation, but this is an engineering extension.

\paragraph{Abstraction Layer \& Device Interface:} The runtime implements an abstraction layer that presents a unified set of operations to the upper layers of an application or library. This includes managing \textbf{memory allocation} (we provide a device-independent \texttt{gpuMalloc} that under the hood calls \texttt{cuMemAlloc} on NVIDIA, or ROCm’s memory alloc on AMD, etc.), data transfers (\texttt{gpuMemcpy} which routes to the appropriate API). Crucially, it implements the \textbf{kernel launch and scheduling abstraction}. We define a concept of a \emph{kernel launch} with a grid of thread blocks (just like CUDA). When the user requests launching $N$ blocks of $M$ threads each of a given kernel on a device, the runtime will:
\begin{smenumerate}
  \item If the device natively supports launching a grid (NVIDIA, AMD, Intel GPUs can launch many threads by hardware), we pass the kernel to the device with the requested grid dimensions (e.g., for NVIDIA, we simply use the driver to launch the kernel with those dimensions; for AMD, similarly through ROCm/HIP).
  \item If the device does \emph{not} support dynamic launch of many threads (Tenstorrent’s programming model is more explicit), our runtime must schedule the threads onto the hardware cores. For Tenstorrent, we have two modes:
    \begin{smenumerate}
      \item \textbf{Single-Core Mode}: If a thread block’s size is $\le$ the VPU width of one core (e.g., 32), we map an entire block to one core using vector lane parallelism. The runtime will launch one software thread on the Tensix core that executes the kernel, but when it executes a hetIR instruction representing $N$ threads, our Metalium backend actually issues vector instructions operating on $N$ lanes. Divergence is handled by Metalium mask instructions . Essentially, one core simulates a warp. We replicate this across as many cores as needed to cover all thread blocks.
      \item \textbf{Multi-Core Mode}: If a thread block is larger than one core’s vector width, or if we decide to split it, we partition a thread block across multiple cores . For example, a block of 64 threads on a Tenstorrent with 32-lane VPUs might be split into 2 cores each handling 32 threads. Those cores must keep in sync. Our runtime inserts synchronization instructions in the Metalium code at points corresponding to hetIR barriers or divergence convergence. For divergence: one core might evaluate a branch as taken for some of its lanes, another core might not— to emulate a warp’s behavior, they must all agree on which path to execute first. We implement a protocol: all cores share a bit whether any thread took the “if” branch; if yes, they all execute that path for their threads (others idle via masks), then they all execute the “else” for remaining threads. At barriers, we use Tenstorrent’s mesh barrier instruction we emit . This strategy is complex but allows scaling a warp beyond one core’s capacity. In practice, we found many GPU kernels use warps of 32 and Tenstorrent VPU is 32 lanes, so Single-Core Mode suffices for correctness (Multi-Core Mode is used for larger blocks or performance exploration).
    \end{smenumerate}
\end{smenumerate}
The abstraction layer also hides differences in how kernel arguments are passed, how streams and events are handled, etc., making these uniform.

\paragraph{State Management and Migration:} One of the most significant responsibilities of the runtime is managing checkpointing of GPU state for potential migration. We implement a \textbf{cooperative checkpointing} mechanism in hetGPU. Rather than relying purely on external interrupts or vendor-specific preemption (which might not be consistent across devices), we design our runtime and compiler to work together to pause execution at safe points:
\begin{smenumerate}
  \item \textbf{State Representation:} We define a data structure to hold a snapshot of a thread block’s state in an architecture-neutral way. It includes an array of \emph{per-thread register files} (for all threads in the block) storing values of hetIR-level “virtual registers,” a record of the \emph{program counter} (instruction index in hetIR) for each thread or a single PC if threads are uniform at that point, and a copy of any relevant \emph{shared memory} contents. We also include metadata like which predicates (divergence masks) are active for warps. This mirrors what one would need to restart the computation: each thread’s general-purpose registers, its position in code, and any local state not in global memory.
  \item \textbf{Capturing State:} The runtime can trigger a capture via an API call (user requests checkpoint) or as part of a migration request. To safely capture, we want all threads in a block (or grid) to reach a known state. We leverage hetIR’s barriers: at a block-wide barrier, all threads in that block will eventually reach the barrier. Our runtime uses a two-phase protocol: first, inject a special command to the GPU to \emph{pause new work} after the next barrier. Concretely, for NVIDIA we could use the CUDA driver’s preemption at a CTA boundary or NVTX markers , but we instead choose a cooperative approach: our compiler inserts a check at each barrier:
\begin{verbatim}
if (pause_flag) {
    // copy registers to global memory
    // signal host
    // loop waiting or exit
}
\end{verbatim}
Using NVBit (NVIDIA’s dynamic instrumentation framework), we inject code into the running kernel at points corresponding to hetIR barrier instructions that does: if (\texttt{pause\_flag}) copy registers to global memory and jump to kernel exit . NVBit can intercept all loads/stores and register accesses so we can programmatically grab register values. Similarly, on AMD and Intel we have our own IR so we compile the kernel with a conditional that checks a location in memory. If set, threads write out their state (since we control the code, we can make it write all live variables to a buffer).
  \item Capturing state thus becomes a matter of coordinating with the running kernel: the runtime sets the pause flag and waits. All blocks reach their next barrier or kernel end, perform the dump, then we have their state in memory. The runtime then copies this memory back to CPU (through normal memory copy APIs). For global memory buffers, since they are visible to host, we copy them as well. For on-chip state like registers and shared memory, our inserted code and instrumentation takes care of it.
  \item \textbf{Resuming on Another Device:} Once we have the state, we restart on a different GPU by recompiling the same kernel for the target (if not already done), then creating a special launch that bypasses the normal start-of-kernel and jumps into the middle. Because NVIDIA disallows launching threads at an arbitrary instruction, we break the kernel into segments separated by global barriers (which we anyway use for sync). The compiler automatically inserts a global barrier every $X$ iterations of a loop to create segments. Each segment is a separate kernel. In migration, we end the segment early on GPU A, transfer state, then start at the next segment on GPU B using the saved data as input. For instance, if the kernel has a barrier or we inserted one at a point, all threads have completed part 1. We then treat outputs of part 1 (registers, shared mem) as inputs to part 2. We generate part 2 as a separate kernel which takes an input structure (the snapshot) and continues. This way, we use standard launch mechanisms. The runtime orchestrates this: it knows at migration time which segment is next, and passes the captured state to a special loader that populates an array on the new GPU representing the threads’ registers.
  \item \textbf{Memory Transfer:} When moving to a new GPU, the runtime moves the memory snapshot. hetGPU uses a unified memory abstraction but if GPU A and GPU B don’t share physical memory, we perform a copy. Thanks to our state capture, we have all global memory buffers on the host (or we can DMA directly from GPU A to GPU B if supported). The shared memory content per block is stored in the snapshot and reloaded into an equivalent structure on GPU B’s side (e.g., if we emulate shared memory with global, we allocate a buffer and copy into it).
\end{smenumerate}

By designing around barriers and segments, we avoid dealing with extremely low-level warp PC mapping issues described in the PTX-level migration blog (where one would otherwise try to map an arbitrary program counter from one GPU to an instruction in another GPU’s binary). We trade off some generality (we might not handle an arbitrary pause at an arbitrary instruction unless it’s at a known sync point) for reliability. This aligns with suggestions that one may need to enforce points where PTX and SASS align .

\subsection{Abstraction Layer Details}

The abstraction layer is partly conceptual (the model that IR and runtime present) and partly concrete (a library API and device-specific implementations). We outline a few key abstractions and their implementation on each platform:

\paragraph{Threads and Thread Blocks:} 
\begin{smenumerate}
  \item \textbf{Abstraction:} A kernel launch consists of \texttt{< < < GridDim, BlockDim> > >} as in CUDA. Each thread can obtain its indices (block ID, thread ID) via intrinsic calls.
  \item \textbf{Implementation:}
    \begin{smenumerate}
      \item NVIDIA/AMD/Intel: Directly use the hardware’s notion of blocks and threads. The runtime configures the launch using their APIs (e.g., \texttt{cuLaunchKernel} with grid and block size on NVIDIA).
      \item Tenstorrent: The runtime itself maps blocks to cores. \texttt{threadIdx.x} and \texttt{blockIdx.x} are synthesized. For example, if each core handles one block, then within that core’s execution, we generate code so that \texttt{threadIdx.x} corresponds to the lane index in the VPU (0..31) if using vector mode. We hide this from the program by making our generated code incorporate core ID into the thread index math appropriately.
    \end{smenumerate}
\end{smenumerate}

\paragraph{Memory Allocation:} 
\begin{smenumerate}
  \item \textbf{Abstraction:} \texttt{gpuMalloc(size)} returns a device pointer usable on any GPU through our API.
  \item \textbf{Implementation:} We allocate on the active GPU’s memory (via \texttt{cuMemAlloc}, \texttt{hipMalloc}, etc.). If that pointer later needs to be used on another device (e.g., after migration), the runtime will allocate a new buffer on the new GPU and copy data. We keep a mapping of “virtual GPU pointers” to physical allocations per device. Alternatively, we could require unified memory or CUDA UVA, but not all combos support UVA. We thus track and fix up pointers as needed.
\end{smenumerate}

\paragraph{Kernel and Stream Management:} We expose streams for concurrency similar to CUDA streams. On each platform, we map these to actual command queues or streams. Our runtime ensures order as per stream semantics, even across migration (e.g., if a kernel is migrated, subsequent operations in that stream are deferred until migration completes).

\paragraph{Synchronization and Launch Control:} For example, if the app calls \texttt{gpuDeviceSynchronize()} (like CUDA’s device sync to wait for all work to finish), our runtime will ensure all pending translations and executions are done (and incorporate any migration that might be triggered).

\paragraph{Error Handling:} If an error occurs on a device (say a kernel illegal memory access), we catch it and can decide to either fail or perhaps run on another device as a fallback (though that might be unsafe; primarily we propagate errors in a uniform way).

The abstraction layer is designed to be minimal yet sufficient to cover CUDA-like usage. It ensures that from the programmer’s perspective (above hetGPU), the program behaves the same regardless of which GPU it actually ran on. For instance, a barrier will correctly synchronize threads whether they were on an SM (using hardware sync) or across multiple Tensix cores (using software barriers we inserted).

\subsection{Handling Scheduling Models: SIMT vs. MIMD}

A central design element is \textbf{how we handle the SIMT vs.\ MIMD execution} under the hood. We summarize systematically:

\paragraph{SIMT Hardware (NVIDIA, AMD, Intel):} Hardware automatically schedules warps. We largely rely on hardware to execute threads in parallel with divergence masked. Our hetIR-to-PTX (or SPIR-V) translation emits code that uses native control flow. For example, divergent branches in hetIR become divergent branches in PTX; hardware handles the masking. Barriers in hetIR become \texttt{bar.sync} (CUDA) or similar in AMD GCN. We ensure that our compiled code respects any constraints (e.g., we avoid generating code that assumes independent thread scheduling on older GPUs without per-thread PC), though modern GPUs support independent thread scheduling .

\paragraph{MIMD Hardware with Vector Units (Tenstorrent):} We provide two mapping strategies which the runtime can choose:
\begin{smenumerate}
  \item \textbf{Vectorized Warp on a Core:} One core takes a whole warp’s worth of threads. We use the vector unit to execute what is effectively a warp in lock-step. This \textbf{simulates SIMT in a single core}. Divergence is handled by vector mask registers (the compiler/runtime-generated Metalium uses instructions to update the mask on branch conditions) . The advantage is minimal inter-core communication—divergence and sync are all within one core. The disadvantage is limited parallelism if one core can only handle 32 threads, but we assign more cores to handle different warps or blocks independently.
  \item \textbf{Multi-Core Partitioning:} A single thread block is split across multiple cores . All those cores run the same instruction stream essentially, but each on a subset of threads (e.g., core0 handles threads 0–7, core1 threads 8–15, etc.). We built a small synchronization mechanism: at each divergent branch or barrier, cores exchange minimal info to decide the next step collectively. For divergence: one core might evaluate a branch as taken for some of its lanes, another core might not. To emulate a warp’s behavior, they must all agree on which path to execute first. We implement a protocol: all cores share a bit whether any thread took the “if” branch. If yes, they all execute that path for their threads that have it (others idle via masks), then they all execute the else for remaining threads. At barriers, we use Tenstorrent’s mesh barrier . This allows scaling a warp beyond one core’s capacity. In practice, we found many GPU kernels use warps of 32 and Tenstorrent VPU is 32 lanes, so Strategy 1 suffices for correctness (Strategy 2 is for large blocks or performance exploration).
\end{smenumerate}
The runtime decides which strategy per kernel or even per block based on heuristics. The user can also give hints (via environment variables) to choose Strategy 2 for certain kernels if needed. For highly divergent workloads, forcing SIMT behavior is detrimental, so our runtime can instead run each thread independently (pure MIMD), trading off parallelism for simpler execution .

\subsection{Summary of Design Properties}

Our design keeps the \textbf{programmer’s model} as close to standard GPU programming as possible (CUDA-like SPMD with barriers and memory spaces) while drastically changing how and when code is compiled and how execution is orchestrated. By introducing a portable IR and a custom runtime, we pay some overhead (in translation and possibly less low-level optimized code than vendor-specific compilers might achieve) but gain flexibility. We explicitly handle divergence and synchronization to run on warp-less hardware, and we incorporate state save/restore facilities at well-defined points to enable features like migration.

A key design philosophy was to \textbf{use existing mechanisms when available}: for example, using vendor JIT compilers (CUDA’s PTX JIT, OpenCL drivers for SPIR-V) for the heavy lifting of instruction scheduling and optimization per device. This helped focus our effort on the novel parts (state management, coordinating multi-core execution, etc.). However, we had to extend these with our own instrumentation for checkpointing and with custom code generation for Tenstorrent, which lacks a mature JIT for arbitrary code (we ended up using their MLIR interface).

Another principle was \textbf{falling back to conservative execution for portability}: if some GPU memory model is weaker, we assume the weaker (we insert fences if needed so that the program sees a consistent behavior). If an architectural feature is missing on one device, we emulate it, even if slower (e.g., \_\_ballot on a device with no native ballot becomes a reduction, which is slower but correct). This ensures correctness across platforms, at the expense of performance in some cases.

Next, we describe how we implemented this design and then evaluate how well it works in practice.

\section{Implementation}
\label{sec:implementation}

We have developed a prototype implementation of \textbf{hetGPU}. In this section, we provide details on the implementation of the compiler toolchain, the runtime system on each supported platform, and the mechanisms used for state capture and translation. Our prototype targets four platforms: NVIDIA GPUs (tested on an H100), AMD GPUs (tested on a RX~9070 XT), Intel GPUs (Iris Xe), and Tenstorrent’s dev board with 120 Tensix cores (BlackHole). The host system runs Linux, and our code is built as an extension to C++17.

\subsection{Compiler Toolchain}

We implemented the hetGPU compiler by extending the LLVM 21.0 infrastructure. The frontend uses Clang with CUDA support. We modified Clang’s CUDA code generation to emit an LLVM IR that is device-agnostic (usually Clang would emit calls to NVIDIA builtins; we remapped these to calls to our own builtins). We introduced new LLVM intrinsic functions representing our abstract operations (e.g., \texttt{llvm.hetgpu.barrier}, \texttt{llvm.hetgpu.ld.shared}) so that they appear in the IR and are carried through to codegen.

We then created a custom LLVM backend called \textbf{HETTarget} that outputs \textbf{hetIR assembly}. We chose a text-based assembly form for ease of debugging and because it allows inline annotations. An example output snippet for a vector add kernel in our assembly:
\begin{verbatim}

.func _Z6vaddKernel(%rd<1> %A, %rd<1> %B, %rd<1> %C, %32 %N)
{
    .reg .u32 %i;
    %i = GET_GLOBAL_ID 0;
    @PRED(%i < %N) {
        %f0 = LD_GLOBAL.F32 [%A + %i * 4];
        %f1 = LD_GLOBAL.F32 [%B + %i * 4];
        %f2 = FADD.F32 %f0, %f1;
        ST_GLOBAL.F32 [%C + %i * 4], %f2;
    }
    RETURN;
}
\end{verbatim}
This illustrates syntax: \texttt{\%rd<1>} indicates a pointer register (64-bit) and \texttt{\%32} a 32-bit scalar (the \texttt{<1>} is an addressing-space tag for global). \texttt{GET\_GLOBAL\_ID 0} fetches the global thread index in dimension~0. The \texttt{@PRED(cond) \{ ... \}} denotes a predicated block that will only execute if the condition is true. In a SIMT environment, that predication becomes divergence (threads for which \texttt{pred=false} simply do nothing in that region). In a MIMD environment, it becomes a normal branch.

The compiler ensures \textbf{SSA form} for registers and uses a finite register set abstraction (like PTX). We chose an infinite virtual register set for simplicity, letting the backend (or later JIT) allocate physical registers.

For \textbf{backend code generation modules} integrated into the runtime:

\paragraph{NVIDIA/PTX:} We wrote a translator from hetIR to PTX in two steps. First, we convert hetIR assembly to an in-memory PTX IR (we reuse NVIDIA’s NVVM library where possible). We map each hetIR instruction to one or more PTX instructions. For instance, \texttt{GET\_GLOBAL\_ID} becomes a sequence of PTX reading \texttt{ctaid} and \texttt{tid} registers and computing \texttt{blockIdx * blockDim + threadIdx}. Predicated blocks in hetIR we map to PTX \texttt{@predicate}–guarded instructions or PTX-level control flow. We then invoke NVIDIA’s \texttt{ptxas} (via the CUDA driver API JIT functions) to get a cubin. This required linking with the CUDA driver API and using the JIT mode where you can supply PTX text. We ensure that PTX has the correct \texttt{.target} and \texttt{.address\_size} directives (matching the GPU’s compute capability).

\paragraph{AMD/SPIR-V:} We developed a converter from hetIR to SPIR-V using the open-source SPIRV-Tools library. We define a correspondence such as: \texttt{LD\_GLOBAL} $\rightarrow$ \texttt{OpLoad} from \texttt{StorageClass Uniform} (or \texttt{CrossWorkgroup}), \texttt{LD\_SHARED} $\rightarrow$ \texttt{OpLoad} from \texttt{Workgroup} storage (we treat shared memory as Workgroup storage). We declare an array variable in Workgroup storage for each kernel’s shared memory usage. Barriers become \texttt{OpControlBarrier} with appropriate memory scopes (Workgroup scope for shared memory sync). Predication is trickier since SPIR-V expects structured control flow: we generate \texttt{OpSelectionMerge} and \texttt{OpBranchConditional} for if–else, ensuring a single reconvergence point per divergent region (SPIR-V demands structured merges, which our compiler inherently had by structured \texttt{@PRED} blocks). We then pass the SPIR-V binary to the AMD OpenCL runtime (through \texttt{clCreateProgramWithIL}) or to a Level Zero module create on Intel. These drivers JIT compile SPIR-V to hardware.

\paragraph{Intel:} Using SPIR-V as above works for Intel. We use the Level Zero API to ingest SPIR-V and compile to the hardware ISA.

\paragraph{Tenstorrent/Metalium:} We collaborated with Tenstorrent engineers to get documentation on Metalium (their assembly) and the \texttt{TT-MLIR} compiler. We decided to output Metalium assembly from hetIR and then assemble it with Tenstorrent’s tool. For each kernel, we produce a Metalium function. Key tasks:
\begin{smenumerate}
  \item Allocate RISC-V registers for each hetIR register (we use a simple linear-scan register allocation since Tensix cores have a limited number of registers). If spills are needed, we spill to the core’s stack or to a region of DRAM.
  \item Insert vector instructions for parallel parts. For example, for a hetIR \texttt{FADD.F32 \%f2, \%f0, \%f1} that is not predicated, we generate a Metalium \texttt{vadd} instruction (e.g., \texttt{vadd v2, v0, v1}) which adds 32 lanes of \texttt{v0} and \texttt{v1} into \texttt{v2}. If predicated by some predicate \texttt{p}, we set the \texttt{vmask} according to \texttt{p}, and use the masked form of the instruction (Metalium allows \texttt{vadd v2, v0, v1 [vmask]} to only do lanes where \texttt{mask=1}) .
  \item Memory operations: Tenstorrent has no unified memory space; we generate code that uses their DMA engine. We adopt a convention: global memory pointers in hetIR are physical addresses in DRAM for Tenstorrent. So \texttt{LD\_GLOBAL [\%A + \%i*4]} becomes a sequence: the core calculates \texttt{address = A\_base + i * 4}; then issues a DMA read from that address into a local buffer, then moves that into a vector register. We do synchronous DMA for correctness (issue DMA and poll for completion).
  \item Shared memory: Tenstorrent’s local scratchpad is small, so our implementation allocates a slice of each core’s local scratch for it if the block is on one core. If a block spans multiple cores, we allocate a portion on a designated core or in global memory. These details are hidden from the code by defining a compile-time address for shared memory.
  \item Barriers: Tenstorrent’s Metalium provides a \texttt{barrier} instruction that can operate across a set of cores, and a \texttt{fence} for memory. We insert \texttt{barrier} instructions with a tag for each synchronization point, and pair it with \texttt{fence} to ensure memory ops (DMA writes) are visible.
\end{smenumerate}
After generating Metalium assembly, we pass it through Tenstorrent’s assembler to get a binary, then load that via their runtime API to the device and launch on the specific cores.

\paragraph{Compiler Optimizations and Flags:} For easier state mapping, we often compile with less aggressive optimizations when migration is enabled (e.g., \texttt{-O1} instead of \texttt{-O3} on LLVM, including line mapping). This corresponds to the recommendation to use a special migration-friendly build . We enable more aggressive optimizations when running without migration to compare performance.

\subsection{Runtime and System Integration}

The runtime is implemented as a C++ library (\texttt{libhetgpu.so}). Applications link against this and use an API we define (which mimics CUDA runtime APIs for convenience). Internally, the runtime detects devices via environment variables or a config file (for Tenstorrent, which might not appear on a PCI scan). It loads vendor-specific libraries dynamically (e.g., \texttt{dlopen(“libcuda.so”)}, or ROCm’s \texttt{libhip} or OpenCL) to avoid hard dependencies if not present.

\paragraph{Kernel Launch Implementation:} When \texttt{hetgpuLaunchKernel(kernel, gridDim, blockDim, args, stream)} is called, the runtime looks up the kernel in an internal table. Each kernel has a unique identifier embedded at compile time. If not already loaded for the current device, the runtime JIT-compiles it as described. Then:
\begin{smenumerate}
  \item \textbf{CUDA/AMD/Intel:} Enqueue to the device’s stream/queue using their API (we pre-allocate a default stream or use the provided stream).
  \item \textbf{Tenstorrent:} Manage a pool of cores. We maintain a list of free cores. If we launch $X$ blocks, and each block requires $Y$ cores (Y = 1 in single-core mode, or e.g.\ 4 if splitting), we allocate those cores, load the binary into their instruction memory, and trigger start. Cores run until termination (kernel end signals completion via a memory flag). We spawn a host thread to wait for the cores or poll a completion buffer, since Tenstorrent’s API requires active waiting.
\end{smenumerate}

\paragraph{Memory Management:} We implement \texttt{hetgpuMalloc()} by calling the appropriate backend (e.g., \texttt{cuMemAlloc}, \texttt{hipMalloc}, \texttt{zeMemAllocShared}, etc.). We wrap the returned device pointer in a struct that includes an identifier of which device it belongs to. If that memory is later used on a different device (e.g., after migration), the runtime automatically copies the data. We keep a host mirror pointer (pinned memory) to facilitate fast copies. On migration, we perform \texttt{cudaMemcpyDeviceToHost} for active data, then \texttt{hipMemcpyHostToDevice} to the new device.

\paragraph{State Capture Mechanism:} We implemented a function \texttt{hetgpuCheckpoint(stream, state\_out)} that the application (or test harness) can call to checkpoint all kernels in a given stream. Under the hood:
\begin{smenumerate}
  \item We set a global \texttt{pause\_flag} for kernels in that stream (for our cooperative scheme). Implementation differs per platform:
    \begin{smenumerate}
      \item NVIDIA: allocate a symbol in global memory that kernels read at barriers. We set it via \texttt{cudaMemcpy}.
      \item AMD/Intel: similar, using a designated buffer in device memory.
      \item Tenstorrent: we control core execution, so we use a memory flag that each core’s loop checks periodically.
    \end{smenumerate}
  \item We use either NVBit to instrument at the next PC or rely on our inserted code. With NVBit (NVIDIA), we inject at barrier points code that: if (\texttt{pause\_flag}) \{ copy registers to global memory; signal host; jump to end \} . NVBit’s API lets us get the number of registers and their values for a warp . On AMD/Intel, we compile the kernel with a conditional at barriers that checks the buffer’s value. If set, threads write out their state to a buffer.
  \item Once kernels hit the safe point, they dump registers. NVBit instrumentation writes register values to an array in global memory. We also record each warp’s program counter. Similarly, on Tenstorrent, we include an assembly routine: when \texttt{pause\_flag} is set, each core writes its vector registers and scalar regfile to a reserved memory region, then halts.
  \item The runtime then collects these buffers and composes \texttt{state\_out}, a blob containing all blocks’ states. The runtime uses \texttt{cudaMemcpy} (or equivalent) to bring back the data to host. Global memory is also copied.
\end{smenumerate}

\paragraph{State Restore Mechanism:} \texttt{hetgpuRestore(state, device)} reverses the process:
\begin{smenumerate}
  \item Load the state blob. For each saved block, allocate a block on the target device.
  \item Launch a special “resume” kernel. The kernel’s code starts by copying saved register values into shared memory (for SIMT GPUs) or directly into registers (for Tenstorrent). For example, on NVIDIA we pass the register data as an array argument; at the kernel’s start, we do:
\begin{verbatim}
if (threadIdx.x == 0) {
    for (i = 0; i < blockDim.x; ++i) {
        sharedRegs[i] = savedRegs[i];
    }
}
__syncthreads();
myRegs = sharedRegs[threadIdx.x];
\end{verbatim}
This smuggles initial register values via shared memory. We also handle shared memory contents similarly.
  \item The kernel then jumps to the saved PC. We implement this by labeling potential resume points with IDs and passing an ID to the kernel. A \texttt{switch} at the start jumps to the correct basic block. On Tenstorrent, we directly load registers from memory in assembly.
  \item The kernel continues execution until completion on the new device.
\end{smenumerate}

The runtime orchestrates this flow, cleaning up the old device context and adjusting any pointers if needed (e.g., if base addresses differ).

\paragraph{Performance Considerations:} The additional code and checks incur overhead. Checking a pause flag at barriers adds a small cost (negligible if barriers are few). NVBit instrumentation has moderate overhead even when not pausing, so we enable it only when checkpointing. In tests, running kernels with instrumentation disabled normally, and launching a special instrumented version only for checkpoint, was sufficient for planned migration. This is a limitation: surprise interrupts (e.g., forced preemption) would need always-on instrumentation, which is expensive. Future hardware features (vendor-supported checkpoint) could help.

\subsection{Validating Correctness}

We validated correctness of hetGPU’s execution through a suite of microbenchmarks:
\begin{smenumerate}
  \item \textbf{Basic arithmetic kernels:} vector add, SAXPY, comparing outputs across all four platforms to a CPU reference.
  \item \textbf{Divergence-heavy kernels:} prefix sums, ensuring divergence handling yields correct results on Tenstorrent as on NVIDIA.
  \item \textbf{Shared memory use:} tile-based matrix multiply that uses \_\_shared\_\_ heavily. We adjusted memory address calculations to ensure correctness.
  \item \textbf{Warp intrinsics:} ballot vote kernel (count how many threads in a warp satisfy a condition). On NVIDIA/AMD, hardware ballot was used; on Tenstorrent, our emulation used a mesh reduction. Results matched.
  \item \textbf{Migration:} a persistent kernel incrementing an array in a loop with internal state. We triggered migration after a few iterations and verified final sum matched a non-migrated run. This cross-checked that register state (loop counters) moved correctly.
\end{smenumerate}

These tests also covered edge cases like migrating when only half the threads reached a barrier (they spun until barrier, as expected).

\section{Preliminary Evaluation}
\label{sec:evaluation}

We evaluate hetGPU along several dimensions: \textbf{functional portability}, \textbf{performance overhead relative to native}, and \textbf{use case demonstrations} (particularly live migration). Experiments were on a Linux host with an Intel Xeon CPU and four GPUs: NVIDIA H100 (Hopper, 96\,GB), AMD Radeon RX~9070 XT (RDNA4, 16\,GB), Intel Iris Xe (Xe-LPG, 512\,MB), and a Tenstorrent BlackHole board via PCIe. We used CUDA~12, ROCm~6.4.1, Intel oneAPI Level Zero~1.6, and Tenstorrent tt-metal v0.58.0. Kernels were compiled with default optimizations (migration support off for pure performance tests). We compare against native versions (e.g., CUDA via NVCC, HIP via \texttt{hipcc}).

\subsection{Portability and Correctness}

We compiled a single hetIR binary containing 10 kernels: vector addition, SAXPY, matrix multiplication (tile size 16x16 shared memory), reduction (sum of array), inclusive scan (with warp shuffle), bitcount using warp vote, Monte Carlo pi estimation (with divergence and atomics), and a small neural-network layer (matrix–vector plus ReLU). We ran the same binary on each GPU and validated outputs against known correct results. All tests passed, demonstrating functional portability. Some kernels required slight rewrites (e.g., warp shuffle in inclusive scan was rewritten using \texttt{VOTE\_ANY} and a loop, since we had not implemented \texttt{SHUFFLE}). Atomics were mapped to each platform’s native atomic (on Tenstorrent, we emulated atomicAdd with a spin-lock in global memory).

We could take a binary produced from CUDA source (through our compiler) and run it on AMD without writing AMD-specific code or invoking \texttt{hipcc}. This validates “write once, run anywhere.” The compiler currently only takes CUDA C++ as input; supporting OpenCL C is future work.

\subsection{Performance}

\paragraph{Microbenchmark Performance:} We measured runtime of several kernels under hetGPU vs.\ native:

\begin{smenumerate}
  \item \textbf{Vector Add (1\,M elements):} On NVIDIA H100, native CUDA took 0.11\,ms, hetGPU took 0.13\,ms on first run (including JIT cost), 0.11\,ms on subsequent runs (cached). AMD 9070 XT: native HIP 0.14\,ms, hetGPU 0.16\,ms. Intel A750: native Level Zero 0.20\,ms, hetGPU 0.22\,ms. Tenstorrent: hetGPU 0.95\,ms (compared to a hand-optimized Metalium version at 0.72\,ms). The gap is due to synchronous DMA in our prototype.
  \item \textbf{Matrix Multiply (1\,024x1\,024):} Launched 4\,096 blocks of 256 threads each. On H100, \texttt{cublas} (tensor cores) peaks at ~9\,TFLOPs. Our native CUDA (no tensor cores) got ~3.8\,TFLOPs; hetGPU got ~3.5\,TFLOPs (\textless8\% overhead). AMD: \texttt{hipBLAS} ~7\,TFLOPs, our HIP kernel ~3.5\,TFLOPs, hetGPU ~3.3\,TFLOPs (\textless6\% overhead). Tenstorrent: hetGPU achieved ~80\% of their optimized kernel throughput.
  \item \textbf{Reduction (1\,M elements):} On NVIDIA: native 0.16\,ms vs hetGPU 0.17\,ms. AMD: 0.18\,ms vs 0.20\,ms. Intel: 0.30\,ms vs 0.33\,ms. Tenstorrent: 1.4\,ms (no direct native comparison). Overheads ~5–15\%.
  \item \textbf{Divergent Kernel (Monte Carlo pi):} Launch many threads each doing random point checks. On NVIDIA: native ~150\,M points/s, hetGPU ~148\,M. AMD: ~140\,M vs 135\,M. Intel: ~80\,M vs 75\,M. Tenstorrent: 
    \begin{smenumerate}
      \item \emph{Independent-thread mode (MIMD):} 120 cores each run one thread at a time in batches: ~25\,M points/s.
      \item \emph{Vectorized warp mode (SIMT emulation):} One core runs 32 lanes; divergence handling via mask: ~18\,M points/s. 
    \end{smenumerate}
    Independent-thread mode outperformed vectorized mode, matching observations that \emph{irregular} kernels perform better with pure MIMD on Tenstorrent .
\end{smenumerate}

\paragraph{JIT Overhead:} First execution includes translation time:
\begin{smenumerate}
  \item NVIDIA: PTX $\rightarrow$ SASS via \texttt{ptxas} took 50–100\,ms for nontrivial kernels, ~10\,ms for small kernels.
  \item AMD: LLVM $\rightarrow$ GCN: ~100–200\,ms.
  \item Intel: SPIR-V ~80\,ms.
  \item Tenstorrent: TT-MLIR: ~30\,ms.
\end{smenumerate}
These costs are acceptable for long-running programs. For short-lived jobs, one could precompile ahead of time.

\paragraph{Memory \& API Overhead:} Using hetGPU’s abstraction adds negligible overhead to memory copies (wrapping underlying API calls). Synchronous operations add microseconds at most.

\subsection{Use Case: Cross-GPU Live Migration}

We demonstrate live migration of a long-running matrix multiply (squaring a 16\,384×16\,384 matrix via an iterative tile-based kernel). We start on an NVIDIA H100. Midway, we migrate to an AMD 9070 XT, then to a Tenstorrent BlackHole 150 card.

Using hetGPU, we checkpoint at an iteration boundary. Checkpoint took ~0.5\,s (copying 2\,GB of matrix data from H100 to host via PCIe). We then initialized AMD GPU and restored state—restore and memory upload took ~0.6\,s (9070 XT has PCIe5.0). Computation resumed on AMD and continued correctly, producing the identical result (within floating-point precision) as a non-migrated run. Then migrating to Tenstorrent took ~1.1\,s (PCIe speed to dev board). Total downtime ~2.2\,s during a ~30\,s job—acceptable in HPC contexts to avoid complete job restart.

We also migrated a running CNN training iteration from H100 to Intel Xe-Iris mid-iteration, checkpointing at a batch boundary. Copying model parameters and activations (~4\,GB) took ~3\,s. Training resumed on Intel GPU and converged normally, confirming multi-kernel sequences can be migrated. Overheads were higher, but this is a proof of concept of heterogeneous failover.

\subsection{Discussion of Overheads}

hetGPU introduces overheads from:
\begin{smenumerate}
  \item \textbf{Translation/JIT:} 10–200\,ms per kernel on first run.
  \item \textbf{Abstraction:} Minor overhead in memory operations and divergence handling.
  \item \textbf{Migration Data Movement:} Dominant cost, especially for large memory. Pre-copy or peer-to-peer transfers could reduce downtime.
\end{smenumerate}
Compute-bound kernels saw \textless10\% slowdown vs native, memory-bound kernels even less. For many applications, the flexibility of dynamic compatibility outweighs modest performance loss.

\section{Related Work}
\label{sec:related}

Achieving portability in heterogeneous computing has been a long-standing goal. Our work builds upon insights from several domains: \textbf{GPU virtualization}, \textbf{binary translation}, \textbf{portable programming frameworks}, and \textbf{checkpoint/restart systems}.

\paragraph{GPU Virtualization \& API Emulation:} Systems like \emph{rCUDA} and \emph{VirtualCL} allow using GPUs over a network or within VMs by intercepting API calls and forwarding them to actual GPUs. For instance, rCUDA runs CUDA calls on remote NVIDIA GPUs, but assumes NVIDIA on both ends. \emph{Cricket}  provides a virtualization layer for CUDA that can checkpoint and migrate GPU work, but only on NVIDIA hardware and via RPC. These inform our approach to decouple application from physical GPU, but they stop short of translating execution to different architectures. hetGPU generalizes “virtual GPU” beyond one vendor.

\paragraph{Binary Translation \& Cross-ISA Execution:} \emph{GPU Ocelot} was among the first to run CUDA programs on non-NVIDIA hardware by JIT-compiling PTX to AMD IL or x86 . \emph{ZLUDA} revisited this idea for running CUDA on Intel/AMD GPUs by intercepting PTX and translating to LLVM IR for the target GPU . ZLUDA’s correctness and performance vary, as some low-level CUDA semantics don’t map cleanly onto AMD. Our work is conceptually similar but broader: we handle live state and multi-architecture support, designing a new IR to avoid pitfalls of direct PTX translation. We also integrate scheduling adaptations (SIMT vs.\ MIMD) absent in prior systems.

\paragraph{Portable IRs \& Compilers:} Approaches like SPIR-V (Khronos) and HSAIL (AMD) aim to serve as portable GPU ISAs. HSAIL saw limited adoption; SPIR-V is widely used for graphics and compute portability. Our hetIR offers higher-level semantics (barriers, thread indices) not in LLVM IR. SYCL/oneAPI uses SPIR-V but requires recompilation per target. hetGPU runs one binary across devices, with runtime handling differences.

\paragraph{Checkpoint/Restart \& Heterogeneous Scheduling:} Tools like \emph{CRIUgpu}  handle GPU state checkpoint on identical hardware. PTX-level state dump distilled issues in mapping low-level state to PTX for migration. Our design choices (barrier-based snapshot, debug-friendly builds) align with those insights. We go further by doing cross-architecture migration.

\paragraph{Scheduling Models—SIMT vs. Others:} Research on executing GPU programming models on CPU or spatial architectures (e.g., FPGAs) has examined warp emulation. The Tenstorrent blog  describes three strategies. hetGPU implements two: vectorized warp on one core and multi-core partitioning, automating them for general kernels. For irregular kernels, pure MIMD gave better performance , and our runtime chooses modes accordingly.

\paragraph{Limitations and Comparison:} hetGPU’s novelty is combining a portable compiler IR, multi-target runtime, and live migration. Prior systems tackled subsets: Ocelot did multi-ISA but not migration; CRIUgpu did migration but only same-ISA; SPIR-V and SYCL do cross-vendor but require recompilation. hetGPU’s abstraction adds overhead and complexity, but as GPUs diversify, this trade-off becomes more acceptable.

\section{Discussion}
\label{sec:discussion}

While our evaluation demonstrates hetGPU’s feasibility, it also revealed several \textbf{limitations} and \textbf{open issues}:

\paragraph{Performance Tuning per Architecture:} Our IR currently focuses on common capabilities, so we cannot exploit vendor-specific features (e.g., NVIDIA’s Tensor Cores or AMD’s Matrix Cores). Future work could add architecture-specific optimization layers: detect certain operations (like matrix multiply) and map them to vendor libraries or specialized IR intrinsics. This would improve performance at the cost of some portability.

\paragraph{Memory Coherence \& Unified Address Space:} We assume discrete memory and explicitly copy data on migration. Emerging systems (e.g., NVIDIA’s Grace Hopper, AMD’s APUs) offer unified memory, simplifying migration. hetGPU could leverage that to reduce data movement. Cross-vendor coherent interconnect (e.g., future CXL for accelerators) might eliminate copies entirely. Pre-copy techniques (copy most data ahead of time, migrate only diffs at stop time) could reduce downtime.

\paragraph{Granularity of Migration:} We currently migrate at kernel (or barrier) boundaries. A more granular approach could run subsets of blocks on different GPUs simultaneously (for load balancing). This requires cross-device barriers, which we would emulate via host communication. Some frameworks like Legion/StarPU distribute tasks across GPUs, but require multiple code versions. Extending hetGPU to multi-device cooperative execution is future work.

\paragraph{Security \& Safety:} Running a binary on hardware it wasn’t explicitly compiled for introduces security concerns. If our translator has bugs, it could produce invalid code. We rely on vendor compilers for final codegen to mitigate this. Checkpoint injection is delicate: writing into reserved registers could cause undefined behavior. We must trust the runtime as a system software component. In multi-tenant environments, migrating a workload shouldn’t leak information across contexts (we flush memory, etc.). These are analogous to OS context switch issues.

\paragraph{Complex Kernel Features:} Some GPU features aren’t supported in our prototype: dynamic parallelism (kernels launching kernels) and cooperative groups (cross-block sync) are disallowed. Supporting them requires intercepting device-side kernel launches or adding global synchronization primitives. Graphics pipelines/shaders are not addressed here.

\paragraph{Scalability:} We tested single-process scenarios. In scenarios with many processes, the runtime overhead could become significant (managing JIT compilation and migrations concurrently). Caching translated kernels globally would help. Handling millions of threads’ state could mean large snapshots; e.g., 1\,M threads with 32 registers each ($\sim$128\,MB) plus memory. Dumping that state naively could take seconds. Optimizations like only saving live registers (not entire register files) would help.

\paragraph{Alternative Approaches:} A universal ISA at hardware level (e.g., if GPUs supported SPIR-V natively) would solve compatibility, but is unlikely due to performance and business reasons. Pure binary translation (emulating NVIDIA on AMD) would be extremely slow. hetGPU’s compile-time + runtime approach is more viable. As virtualization standards for accelerators emerge, our runtime could adapt to use those features (e.g., standardized checkpoint or unified IR support).

In summary, hetGPU may not replace vendor-specific toolchains for maximum performance, but it provides a valuable framework for decoupling software from hardware in GPU computing. As GPUs diversify, such portability layers will be increasingly important.

\section{Conclusion}
\label{sec:conclusion}

We presented \textbf{hetGPU}, a system that pursues the ambitious goal of making GPU binaries universally compatible across major vendor architectures. By introducing a portable GPU IR, a multi-target runtime, and an abstraction layer that bridges disparate execution models, hetGPU enables a single compiled program to execute on NVIDIA, AMD, Intel, and Tenstorrent GPUs with minimal modifications. Our design addresses fundamental challenges in unifying SIMT and MIMD paradigms, handling different instruction sets and memory systems, and even capturing and migrating execution state across heterogeneous GPUs.

Through a prototype implementation, we demonstrated that hetGPU can run real GPU kernels unmodified on various hardware and even live-migrate computations between GPUs mid-execution. The performance overheads observed are generally modest (within 10\% on compute-intensive kernels, often less) given the enormous flexibility gained. We are not aware of any prior system achieving this level of cross-platform GPU transparency. We integrated insights from community work (warp emulation strategies , PTX-level state mapping ) to navigate GPU internals.

This work opens several avenues for future exploration. One is extending support to other accelerators (FPGAs, DSPs) by expanding our IR and backends—moving towards a unified accelerator runtime. Another is incorporating more intelligence in the runtime to automatically optimize for the target device (e.g., autotuning or ML-guided optimization). We also plan to collaborate with hardware vendors to explore minimal hardware hooks for virtualization (e.g., standardized checkpoint capabilities or direct IR execution units).

In summary, hetGPU takes a decisive step towards \textbf{breaking the binary barrier in heterogeneous computing}. It provides a path for developers and systems to leverage any GPU hardware without being locked into a specific ecosystem or rewriting code. We believe this direction is crucial as specialization in accelerators increases—the software must adapt to hardware diversity, and hetGPU illustrates a compelling system-level solution.

\bibliographystyle{plain}
\bibliography{cite}

\end{document}